\documentclass{mn2e}

\usepackage{epsfig}

\newcommand{\ve}[1]{{\bmath {#1}}}

\newcommand{\pa}[2]{\frac{\partial {#1}}{\partial{#2}}}

\newcommand{\clm}{C^{lm}(r)}
\newcommand{\dlm}{D^{lm}(r)}
\newcommand{\grad}{{\bmath{\nabla}}}
\newcommand{\half}{\frac{1}{2}}
\newcommand{\nr}{N_{\rm{r}}}
\newcommand{\nth}{N_{\theta}}
\newcommand{\nph}{N_{\phi}}
\newcommand{\ijk}{_{\rm{i,j,k}}}
\newcommand{\rjk}{{R_{\rm{j,k}}}}
\newcommand{\ri}{r_{\rm{i}}}
\newcommand{\thj}{\theta_{\rm{j}}}
\newcommand{\phk}{\phi_{\rm{k}}}
\newcommand{\rhoc}{{\rho_{\rm{c}}}}
\newcommand{\rs}{{R_{\rm{s}}}}
\newcommand{\plm}{{P_l}^m}
\newcommand{\plmp}{{P_l}^{m+1}}
\newcommand{\uryu}{Ury$\bar{\rm{u}}$ \& Eriguchi}
\newcommand{\muller}{M$\ddot{\rm{u}}$ller}
\newcommand{\zeus}{{\sevensize{ZEUS-2D}}}
\newcommand{\eq}{\begin{equation}}
\newcommand{\ee}{\end{equation}}
\newcommand{\eqa}{\begin{eqnarray}}
\newcommand{\eea}{\end{eqnarray}}
\newcommand{\cross}{{\bmath \times}}
\newcommand{\dotp}{{\bmath \cdot}}
\newcommand{\ts}{^{t+\frac{\delta t}{2}}}

\date{Accepted 2002 April 22.}

\pubyear{2002}
\pagerange{\pageref{firstpage}--\pageref{lastpage}}

\title[A three-dimensional fluid dynamics code for 
stars in binaries]
{A general, three-dimensional fluid dynamics code for 
stars in binary systems}

\author[M. E. Beer and Ph.\ Podsiadlowski]{Martin
E. Beer\thanks{E-mail: beer@astro.ox.ac.uk} and Philipp Podsiadlowski
\\University of Oxford, Wilkinson Building, Keble Road, Oxford,
OX1 3RH, England}

\begin{document}

\maketitle
\label{firstpage}

\begin{abstract}
We describe the theory and implementation of a three-dimensional fluid 
dynamics code which we have developed for calculating the 
surface geometry and circulation currents in the secondaries of 
interacting binary systems. The main method is based on an 
Eulerian-Lagrangian scheme to solve the advective and force terms
in Euler's equation. Surface normalised spherical polar coordinates are
used to allow the accurate modelling of the surface of the star, as is
necessary when free surfaces and irradiation effects are to be considered. 
The potential and its gradient are expressed as sums of Legendre polynomials,
which allows a very efficient solution of Poisson's equation. The basic
solution scheme, based on operator splitting, is outlined, and standard
numerical tests are presented.
\end{abstract}


\section{Introduction}
Irradiation of the secondaries in X-ray binaries can dramatically
change their appearance and their internal structure. The irradiation
pressure force can lead to significant distortions of the surface
(Phillips \& Podsiadlowski 2002), while irradiation-driven circulation
currents can transport significant amounts of energy to the unirradiated
side. There is ample observational evidence
for the existence of such circulation currents: e.g. in HZ
Her/Her X-1 (Kippenhahn \& Thomas 1979; Schandl, Meyer-Hofmeister 
\& Meyer 1997),
in cataclysmic variables  (Davey \& Smith 1992), in Nova Sco
during outburst (Shahbaz et al.\ 2000) and in Cyg X-2 (J. Casares \&
P.A. Charles 1999; private communication), there is clear evidence that
a substantial amount of X-ray heated material can move beyond the
X-ray horizon. Moreover, persistent residuals in the observed
radial-velocity curves in X-ray binaries, e.g. in Nova Sco during
outburst (Orosz \& Bailyn 1997) and Vela X-1 (Barziv et al.\ 2001),
may provide direct evidence for circulation.

Modelling of the irradiation-induced circulation in binaries is 
difficult due the three-dimensional nature of the problem. It requires the
simultaneous solution of the shape of the irradiated star and the
circulation and a proper treatment of the surface boundary conditions.

\begin{figure*}
\center
\epsfig{file=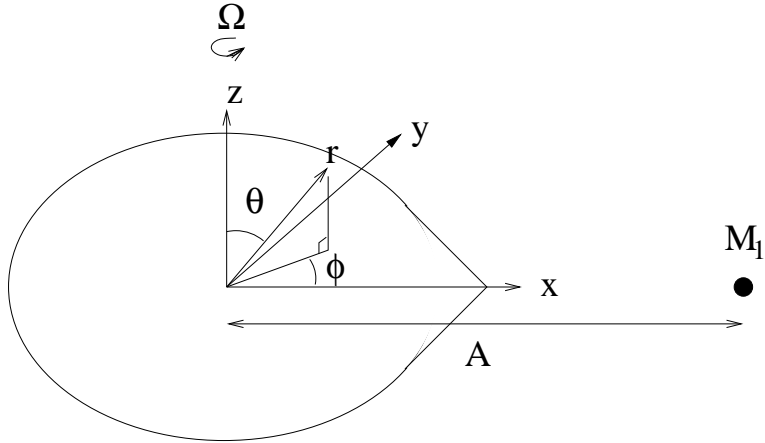,width=6cm,angle=-90}
\caption{Schematic diagram defining the coordinate systems
($r,~\theta,~\phi$) and ($x,~y,~z$) relative to the 
system geometry and other binary parameters ($A$: orbital separation,
$M_1$: mass of the primary, $\ve{\Omega}$: angular velocity of the secondary).}
\label{sphpolars}
\end{figure*}

Attempts to model circulation in irradiated secondaries have been made
by various authors in the past, initially using perturbative methods
either in planar geometry (Kippenhahn \& Thomas 1979; K\i rb\i y\i k
\& Smith 1976) or spherical geometry (K\i rb\i y\i k 1982), and more
recently non-perturbatively using smooth-particle hydrodynamics
(Martin \& Davey 1995).  None of these investigations to-date,
however, are self-consistent, neither allowing for changes in the
surface geometry nor including radiative surface stresses.

On the other hand, circulation currents have been extensively modelled
in multi-dimensions in the context of modelling the circulation in the
ocean and tidal flows on Earth,
where over the years efficient methods have been
developed to treat circulation realistically and in a numerically
efficient way.  A commonly used method is based on an
Eulerian-Lagrangian scheme (a description of which may be found in Lu
\& Wai 1998). This method is similar to the upwind method but more
physically sound in the physical treatment of advective terms and has
been shown to be unconditionally stable (Casulli 1990; Casulli \& Cheng 1992).

The purpose of this paper is to present the philosophy and the details of a
fairly general three-dimensional fluid dynamics code which we have specially
developed for treating the secondaries in interacting binaries,
in particular under the influence of external irradiation. The main
method is an application of the Eulerian-Lagrangian method by
Lu \& Wai (1998) which we have modified for our application, drawing
also on the results of related work by Ury\=u \& Eriguchi (1995, 1996, 1998), 
\muller~\& Steinmetz (1995) and using methods developed in the context
of geophysical fluids (see e.g. Pedlosky 1987). At present our code
is still somewhat simplified since we do not include the thermodynamic 
equation, necessitating the use of a polytropic equation of state.
In Appendix~B, we describe
how we plan to generalize the code in the future.
In a subsequent paper, we will apply the code first to rotation
in the standard Roche problems, and then to study 
irradiation-driven flows in X-ray binaries, considering both the effects
of heating and external radiation pressure, and of radiative
surface stresses.

The outline of the paper is as follows. In Section~\ref{coordinate} we
describe the transformations and the usefulness of surface fitting
coordinates, Section~\ref{basiceq} describes the basic equations and
the dimensionless variables used.  The theory and implementation of
the calculation of the gradient of the gravitational potential is
given in Section~\ref{potential}, including estimates of its
accuracy. The general solution method is described in
Section~\ref{solmethod}. Finally,
Sections~\ref{advect}~and~\ref{comparison} present standard numerical
tests of the code, advection tests and Maclaurin spheroids.

\section{Coordinate system} \label{coordinate}

Figure~\ref{sphpolars} defines the adopted coordinate system, 
standard spherical polar coordinates centered on the center of mass
of the secondary, where
the directions of the axes are given by the unit vectors 
($\ve{\hat{r}},~\ve{\hat{\theta}},~\ve{\hat{\phi}}$).

\subsection{Surface normalised coordinates} \label{surface}
For our main applications it is important to model the stellar surface
accurately and to allow it to adjust freely to satisfy whatever
surface boundary conditions are applied.  If the outer edge of the
coordinate grid did not coincide with the surface, it would be
non-trivial to calculate surface stresses and derivatives along the
surface accurately.  These difficulties can be avoided by using a grid
whose boundary is defined by the stellar surface. To achieve this, we
follow \uryu~(1996) and  transform our basic equations (see 
Section~\ref{basiceq}) using spherical surface fitting coordinates, i.e.
\eq
r^*=\frac{r}{R(\theta , \phi )}~, \quad\theta ^* = \theta ~, \quad\phi
^* = \phi ~,
\ee
where $R(\theta , \phi )$ is the radius of the star in the direction
 $(\theta ,~\phi)$. With this definition, $r^*$ is restricted to the range
\eq
0 \le r^* \le 1 ~.
\ee
This means that the grid has to adapt continually as the stellar surface
changes and that derivatives are transformed according to:
\eq R(\theta , \phi ) \rightarrow R(\theta ^*, \phi ^*) ~,
\label{surfeqbeg} \ee 
\eq \pa{}{r} \rightarrow \frac{1}{R(\theta , \phi
)}\pa{}{r^*} ~, \ee
\eq \pa{}{\theta} \rightarrow \pa{}{\theta ^*} -
\frac{r^*}{R(\theta ^*, \phi ^*)} \pa{R(\theta ^*, \phi ^*)}{\theta
^*} \pa{}{r^*} ~, \ee
\eq \pa{}{\phi} \rightarrow \pa{}{\phi ^*} -
\frac{r^*}{R(\theta ^*, \phi ^*)} \pa{R(\theta ^*, \phi ^*)}{\phi ^*}
\pa{}{r^*} ~. \label{surfeqend} \ee
The surface derivatives $\pa{R}{\theta ^*}$ and $\pa{R}{\phi ^*}$ are
calculated using the potential as described in
Appendix~\ref{surfderiv}.

\subsection{Coordinate grid}
\begin{figure}
\begin{center}
\epsfig{file=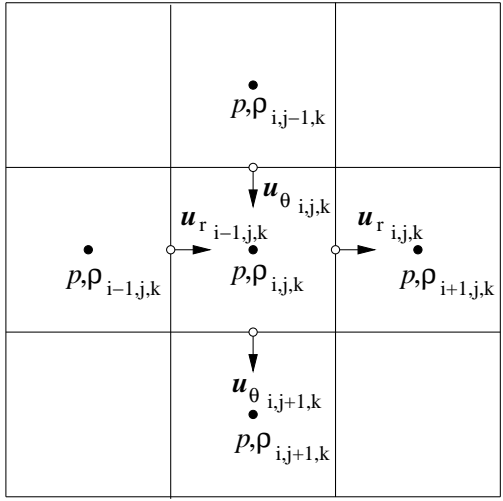,height=7cm,angle=-90}
\end{center}
\caption{Cross-section of the adopted coordinate grid in the $r$ and
$\theta$ directions, indicating where quantities (velocities, $u$,
pressure, $p$, and density, $\rho$, are evaluated.}
\label{gridfig}
\end{figure}

To define the coordinate grid, we split 
the star into equally spaced intervals in $r,~\cos\theta$ and 
$\phi$ with a total of $\nr,~\nth,~\nph$ elements. The equal spacing
in $\cos\theta$ ensures that all grid cells have the same volume. 
The centers of the grid cells in the $\theta$ and $\phi$ directions
are defined by
\eq \cos\thj = 1-\frac{(j-0.5)}{\nth}~, \quad \phk = \frac
{2\pi(k-1)}{\nph} ~, \ee
where we restricted the $\theta$ range assuming even symmetry with respect 
to the xy-plane, 
and the surface normalised radial coordinate becomes
\eq r\ijk = \frac{(i-0.5) R(\thj, \phk)}{\nr} ~, \quad r^*\ijk =
\frac{i-0.5}{\nr} ~. \ee
Velocities are calculated on cell boundaries whilst pressures and
densities are evaluated at the centres of cells. Figure~\ref{gridfig}
shows a cross-section of the grid indicating where
various quantities are evaluated.

This grid spacing acts so that the entire grid adapts when the surface
changes. We adapt this so that if desired the central grid region stays
fixed at its initial position when the surface adjusts. This enables
us to fix the innermost regions of the grid in situations when we are
only interested in the outermost layers. For this purpose we define
$N_{\rm c}$ as the number of radial layers from the center whose
positions are kept fixed.
\eqa r\ijk = \left\{ \begin{array}{ll}
\frac{(i-0.5) R_{\rm in}(\thj, \phk)}{\nr} & \quad {\rm if}~ i \leq
N_{\rm{c}} ~,\\
\rule{0pt}{15pt}\frac{N_{\rm c} R_{\rm in}(\thj, \phk)}{\nr} +
\frac{(i-0.5-N_{\rm c})}{(\nr-N_{\rm c})} \times\\
\hfill \rule{0pt}{12pt}\left [ R(\thj, \phk) -
\frac{N_{\rm c} R_{\rm in}(\thj, \phk)}{\nr} \right ] & \quad {\rm
otherwise} ~.
\end{array} \right .\hfill \eea
If desired it is possible to redefine the radial spacing so that it is
no longer linear but concentrated in a particular region, e.g. near
the surface. The number of grid points, however, is kept fixed during a
calculation. Fixing the innermost region of 
the grid also requires an alteration to equations
(\ref{surfeqbeg}--\ref{surfeqend}), since the term
proportional to the surface curvature has to be replaced by a term
representing the curvature of the grid at the required radius.
\eqa r^*\pa{R}{\theta} \rightarrow \left\{ \begin{array}{ll}
r^*
\pa{R_{\rm in}}{\theta} & \quad {\rm if}~ i \leq
N_{\rm c} ~, \\
\rule{0pt}{15pt}\frac{N_{\rm c}}{N_{\rm r}} \pa{R_{\rm
in}}{\theta} + \frac{i-0.5-N_{\rm c}}{N_{\rm r} - N_{\rm c}} \times\quad\\
\hfill \rule{0pt}{12pt}\left (
\pa{R}{\theta} - \pa{R_{\rm in}}{\theta} \right ) & \quad {\rm
otherwise} ~.
\end{array} \right .\hfill\eea

\section{Basic Equations} \label{basiceq}
The equation of motion relating velocity ($\ve{u}$), pressure ($p$),
density ($\rho$), frictional force ($\ve{\mathcal{F}}$) and potential
($\Phi '$) in a frame rotating about the centre of mass of
the secondary with angular velocity $\ve{\Omega}$ is
\eq \frac{d\ve{u}}{d t} + 2\ve{\Omega}\cross\ve{u} = -\frac{1}{\rho}\grad p
- \grad\Phi '  + \ve{\mathcal{F}} ~, \label{euler} \ee
where
\eq \frac{d\ve{u}}{d t} = \pa{\ve{u}}{t} + (\ve{u} \dotp \grad)\ve{u} ~,\ee
and the potential term is given by
\eq \Phi ' = \Phi + \frac{GM_1r\sin\theta\cos\phi}{A^2} -
\frac{(\ve{\Omega}\cross\ve{r})^2}{2} -\frac{GM_1}{r_2} ~, \ee
where the second term represents a coordinate transformation from the
centre of rotation to the centre of the secondary.
Here $M_1$ is the primary mass, $G$ the gravitational constant,
$A$ the binary separation,
$r_2$ is the distance from the primary to the point
($r,\,\theta,\,\phi$), and the potential of the secondary $\Phi$ is
given by Poisson's equation
\eq
\nabla^2 \Phi = 4\pi\,G\rho~.
\ee
The continuity equation,
\eq \frac{d\rho}{d t} + \rho\grad \dotp \ve{u} = 0 ~, \ee
can be written in integral form, using the divergence theorem, as
\eq \pa{}{t} \int_V \rho\, d V + \int_{d V} \rho\, \ve{u} \dotp \ve{\hat{n}}\, 
d S =
0 ~, \label{conuity} \ee
where $\ve{\hat{n}}$ is a unit surface normal vector. These equations need
to be  solved along with an equation of state, assumed to be a polytropic 
equation of state at present which relates pressure and
density according to
\eq p = K\rho^{\left ( 1+\frac{1}{n}\right ) } ~, \label{polytrope}
\ee
where $K$ and $n$ are the polytropic constant and polytropic index,
respectively.

\subsection{Boundary conditions}

To completely specify the mathematical problem, we also need to 
specify a set of boundary conditions
at the centre and the surface. The central boundary conditions
are unproblematic and are given by
\eq \ve{u} = 0 ~, \quad \pa{u_{\rm{r}}}{r} = 0 ~. \ee
The simplest boundary condition for the surface are the zero-pressure
condition and the conservation of the mass of the secondary, 
\eq p = 0 ~, \quad  M_2 = \mbox{constant} ~,\ee
or, more realistically, a 
pressure condition that assumes an atmosphere in hydrostatic equilibrium,
i.e.
\eq p = \frac{2g}{3\kappa} ~, \ee
where $g$ is the effective surface gravity, and $\kappa$ the photospheric 
value of the opacity (see e.g. Kippenhahn \& Weigert 1994).

In the case of irradiated stars, a special treatment of the surface
layer is required since the irradiation flux is deposited in a thin
(turbulent) surface layer (of order an atmospheric scale height),
which will generally be smaller than our grid size. A separate
atmosphere calculation, which includes the effects of heating and
irradiation pressure, then determines the pressure across the surface
of the star (analogously to the case of normal stars; also see Tout et
al.\ 1989). In addition, the variation of the radiation pressure force
causes a surface stress which drives horizontal motion perpendicular
to the stress and vertical motion in a thin turbulent boundary layer
(an `Ekman' layer), a process known as `Ekman' pumping. This produces
a vertical velocity component in regions where the surface stress
varies.  This process is entirely analogous to the wind-driven
circulation in oceanic circulation systems (see chapter 5 of Pedlosky
1987) and can be treated analogously.

\subsection{Surface adjustment}

At the end of each time step, the surface is adjusted using the current
values of the velocity at the free surface. The surface normal is
given by
\eq {\ve{\hat{n}}} = \ve{\hat{r}} - \frac{1}{r}\pa{R}{\theta} 
\ve{\hat{\theta}} - \frac{1}{r\sin\theta} \pa{R}{\phi} \ve{\hat{\phi}}
~. \ee
The dot product of the surface velocity
with the surface normal gives the velocity component normal to the
surface. Equating this to the component of the surface adjustment in the
radial direction ($\delta r$), which is normal to the surface, yields
\eq \delta r = \frac{\ve{u} \dotp \ve{\hat{n}}}{\ve{\hat{n}} \dotp
\ve{\hat{r}}} ~. \ee
This is added to the current value of $r$ at the end of each time step.
In order to ensure mass conservation, the density of each $\theta,
\phi$ element is scaled by the ratio of the old and new radii cubed
\eq \rho \rightarrow \left ( \frac{r}{r+\delta r} \right ) ^3 \rho
~. \ee
This ensures that the density decreases as the radius increases, conserving
mass in the process. 
If mass is not conserved instabilities can develop at the surface of
the star. For example, consider the case of a contracting
star. As mass leaves the inner radial boundary of the surface layer, no mass
loss/gain occurs through the upper boundary and so the density in
the surface layer drops to zero.
The central density is calculated by extrapolating the
densities in the central regions of the star. It is only used as a
boundary condition in the potential calculation
(Section~\ref{potential}). The extrapolation occurs by extrapolating
in the radial direction for each $\theta$ and $\phi$ direction before
taking the mean of these values.

\subsection{Viscosity}
Artificial viscosity is often added in numerical
simulations to smooth out the flow and to broaden shock fronts.
Stone \& Norman (1992) in their code \zeus~use an artificial pressure
similar to that of von Neumann \& Richtmyer (1950):
\eqa
q =
\left\{
\begin{array}{ll} 
C^2 \rho (\Delta v)^2 & \quad{\rm if}~(\Delta v) < 0 ~, \\
0 & \quad{\rm otherwise} ~,
\end{array}
\right.
\eea
where $\Delta v$ is the change in velocity across a cell and $C =
l/\Delta\rm{x}$. $C$ measures the number of zones over which the
artificial viscosity spreads a shock and is typically chosen to be
$C \sim 3$. This has been generalised to multidimensions by
Tscharnuter \& Winkler (1979) using an artificial viscous pressure
tensor ({\bf\sf Q})
\eqa
{\bf\sf Q} = \left\{ \begin{array}{ll}
l^2 \rho \grad \dotp \ve{u} \left [ {\bf\sf \nabla u } - \frac{1}{3}(\grad
\dotp \ve{u} ) {\bf\sf e} \right ] & \quad {\rm if}~ \grad \dotp \ve{u} < 0 ~, \\
0 & \quad {\rm otherwise} ~,
\end{array}
\right .
\eea
where {\bf\sf e} is the unity tensor, ${\bf\sf \nabla u }$ is the
symmetrized tensor of the velocity field and $l$ is of the order of
the local width of the grid. Following Stone \& Norman (1992), we
neglect the off-diagonal (shear) components of the artificial viscous
pressure tensor. The (nonzero) diagonal elements of the grid are 
\eqa {\bf\sf Q_{11}} = l^2 \rho \grad \dotp \ve{u} \left [ \pa{u_r}{r} -
\frac{1}{3}(\grad \dotp \ve{u} ) \right ] ~, \eea
\eqa {\bf\sf Q_{22}} = l^2 \rho \grad \dotp \ve{u} \left [ \frac{1}{r}
\pa{u_{\theta}}{\theta} + \frac{u_r}{r} - \frac{1}{3}(\grad
\dotp \ve{u} ) \right ] ~, \eea
\eqa {\bf\sf Q_{33}} = l^2 \rho \grad \dotp \ve{u} \bigg
[\frac{1}{r\sin\theta} \pa{u_{\phi}}{\phi} + 
\frac{u_r}{r} + \frac{u_{\theta}}{r} \cot\theta - \nonumber\eea
\eq\phantom{empty} \hfill \frac{1}{3}(\grad \dotp \ve{u} ) \bigg ] ~.~ \ee
The frictional force in the momentum equation (\ref{euler}) is
\eq \ve{\mathcal{F}} = - \frac{1}{\rho} \grad \dotp {\bf\sf Q} ~, \ee
where
\eqa \ve{\hat{r}} . \ve{\mathcal{F}} = - \frac{1}{\rho} \pa{\bf\sf
Q_{11}}{r} - \frac{3{\bf\sf Q_{11}}}{\rho r}  ~,  \nonumber\eea
\eqa \ve{\hat{\theta}} . \ve{\mathcal{F}} = - \frac{1}{\rho r} \pa{\bf\sf
Q_{22}}{\theta} - \frac{\left ( {\bf\sf Q_{11} + 2 Q_{22}}
\right)}{\rho r} \cot\theta ~, \nonumber\eea
\eqa \ve{\hat{\phi}} . \ve{\mathcal{F}} = \frac{1}{\rho r\sin\theta}\pa{\left
( {\bf\sf Q_{11} + Q_{22}} \right)}{\phi} ~, \eea
and where we have used the property ${\bf Tr(\sf Q)} = 0$ to eliminate
${\bf\sf Q_{33}}$. This source of viscosity may be switched on or off
in the code.

Stone \& Norman (1992) also consider an artificial linear viscous pressure,
\eq q = C_1 \rho C_{\rm{a}} \Delta v ~, \ee 
to damp instabilities in stagnant regions of the flow,
where $C_1$ is a constant of order unity and $C_{\rm{a}}$ is the
adiabatic sound speed. This  is calculated separately for each
direction and then added, e.g. in the $r$ direction, as
\eq
\ve{\hat{r}} \dotp \ve{\mathcal{F}} = - \frac{1}{\rho}
\pa{q_{\rm{r}}}{r} ~.
\ee
In our code we include the linear viscous pressure which may be
switched on or off.

\subsection{Dimensionless variables} \label{dimensionless}
In the code we use dimensionless variables defined as:
\eqa \tilde{R} = \frac{R}{\rs} ~, \quad
\tilde{\rho} = \frac{\rho}{\rhoc} ~, \quad
\tilde{\grad} = \rs\grad ~, \quad
\tilde{p} = \frac{p}{G\rhoc^2\rs^2} ~, \quad\nonumber\eea
\eqa
\tilde{\Phi} = \frac{\Phi}{G\rhoc \rs^2} ~,\quad 
\tilde{\ve{u}} = \frac{\ve{u}}{\rs\sqrt{G\rhoc}} ~, \quad
\tilde{\Omega} = \frac{\Omega}{\sqrt{G\rhoc}} ~, \quad
\nonumber\eea
\eqa
\tilde{M} = \frac{M}{\rhoc \rs^3} ~, \quad
\tilde{L_{\rm{x}}} = \frac{L_{\rm{x}}}{G\rs^2 c} ~, \quad
\tilde{t} = t\sqrt{G\rhoc} ~, \quad
\nonumber\eea
\eqa
\tilde{\ve{\mathcal{F}}} = \frac{{\ve{\mathcal{F}}}}{G\rhoc \rs} ~,
\quad
\tilde{\nu} = \frac{\nu}{\rs^2\sqrt{G\rhoc}} ~, \quad
\tilde{\kappa} = \kappa \rhoc \rs ~, \eea
where $\rs$ is the solar radius, $\rhoc$ the initial central
density and $G$ the gravitational constant.
This transforms equations~(\ref{euler}),~(\ref{conuity})~\&~(\ref{polytrope})
to:
\eq \frac{d\tilde{\ve{u}}}{d\tilde{t}} +
2\ve{\tilde{\Omega}}\times\ve{\tilde{u}} = -
\frac{1}{\tilde{\rho}}\tilde{\grad} \tilde{p} -
\tilde{\grad}\tilde{\Phi '} + \ve{\tilde{\mathcal{F}}} ~, \ee

\eq \pa{}{\tilde{t}} \int_{\tilde{V}} \tilde{\rho}\, d\tilde{V} +
\int_{d\tilde{V}} \tilde{\rho} \,\ve{\tilde{u}} \dotp \ve{\hat{n}}
\, d\tilde{S} = 0 ~, \ee

\eq \tilde{p} = \frac{K\tilde{\rho}^{\left ( 1+\frac{1}{n}\right )
}}{G \rs^2\rhoc^{\left ( 1-\frac{1}{n}\right ) }}~. \ee

\section{Potential calculation} \label{potential}
\subsection{Theory}
\muller~\& Steinmetz (1995) developed an efficient algorithm
for solving Poisson's equation which utilizes spherical coordinates
and an expansion into spherical harmonics. This results in an
algorithm which has a computational cost proportional to $(L+1)^2
\nr\nth\nph$ where $L$ is the highest order harmonic considered. The
general solution of Poisson's equation in
spherical harmonics, $Y^{lm}(\theta , \phi )$ (Morse \& Feshbach 1953)
can be written as
\eqa \Phi(r, \theta , \phi ) = -G \sum_{l=0}^{\infty} \frac{4\pi}{2l+1}
\sum_{m=-l}^l Y^{l m}(\theta , \phi) \times\nonumber\eea
\eq \phantom{empty}\hfill\left[\frac{1}{r^{l+1}}\clm+r^l\dlm\right ]
~,\ \label{mullereq}\ee
where
\eq \clm = \int_{4\pi} d\Omega ' Y^{l m*}(\theta ', \phi ') \int_0^r d r' 
 {r'}^{l+2} \rho(r', \theta ', \phi ') \label{cl} ~, \label{eqclm}\ee

\eq \dlm = \int_{4\pi} d\Omega ' Y^{l m*}(\theta ', \phi ')\int_r^{\infty} d r'
 {r'}^{1-l} \rho(r', \theta ', \phi ') \label{dl} ~, \label{eqdlm}\ee

 \eq d\Omega ' = \sin \theta ' d\theta ' d\phi ' ~. \ee
By differentiating this equation we may obtain analytical formulae for
the gradient of the potential in spherical coordinates (as derived
in Appendix~\ref{dplmdth}):
\eqa \pa{\Phi(r, \theta , \phi )}{r} = -G \sum_{l=0}^{\infty}
\frac{4\pi}{2l+1} \sum_{m=-l}^l Y^{l m}(\theta , \phi)
\times\nonumber\eea
\eq\phantom{empty}\hfill
\bigg[\frac{-(l+1)}{r^{l+2}}\clm + l r^{l-1}\dlm \bigg ] ~.~
\label{dpotdr} \ee 
\eqa
\pa{\Phi(r, \theta , \phi )}{\theta} = -G \sum_{l=0}^{\infty}
\frac{4\pi}{2l+1} \sum_{m=-l}^l Y^{l m}(\theta , \phi)
\times\nonumber\eea 
\eq 
\left
[\frac{P_l^{|m+1|}(\cos \theta)}{P_l^{|m|}(\cos \theta )} + 
\frac{m\cos\theta}{\sin\theta} \right ] 
\left [ \frac{1}{r^{l+1}}\clm+r^l\dlm \right ]
~.\ \label{dpotdth}
\ee
\eqa \pa{\Phi(r, \theta , \phi )}{\phi} = -G \sum_{l=0}^{\infty}
\frac{4\pi}{2l+1} \sum_{m=-l}^l i m Y^{l m}(\theta , \phi )
\times\nonumber\eea
\eq\hfill \left
[\frac{1}{r^{l+1}}\clm+r^l\dlm \right ] ~.~ \label{dpotdph} \ee
For odd $l+m$, $\plm(\cos\theta ')$ is an odd function in $\theta '$.
Since the density is an even function w.r.t. $\theta '$ (because of the assumed
symmetry with respect to the $xy$-plane) $\rho(\theta ', \phi ',r')
\sin\theta ' 
\plm(\cos\theta ')$ is an odd function and its integral
from $\theta = 0$ to $\pi$ in equations~%??
(\ref{eqclm}) \& (\ref{eqdlm}) 
is zero.
Hence terms odd in $l+m$ do not contribute to the sum, and we only need
to include values of $m$ from 0 to $l$ and
double the contribution of the positive $m$ term.

In our code we deal with $\Phi '$ rather than $\Phi$,
\eqa \Phi '(r, \theta , \phi ) = \Phi(r, \theta , \phi ) -
\nonumber\eea\eq\hfill 
\frac{(\Omega r\sin \theta )^2}{2} +
\frac{GM_1r\sin\theta\cos\phi}{A^2} - \frac{GM_1}{r_2} ~.\ \ee
This then yields:
\eqa \pa{\Phi '}{r} = \pa{\Phi}{r} - r\Omega ^2\sin ^2\theta +
\nonumber\eea\eq\hfill 
\frac{GM_1\sin\theta\cos\phi}{A^2} + 
\frac{GM_1(r-A\sin\theta\cos\phi
)}{(A^2 + r^2 -2A r\sin\theta\cos\phi)^{\frac{3}{2}}} ~,\ \ee
\eqa 
\pa{\Phi '}{\theta} = \pa{\Phi}{\theta} - (r\Omega
)^2\sin\theta\cos\theta + \nonumber\eea\eq\hfill 
\frac{GM_1r\cos\theta\cos\phi}{A^2} -
\frac{GM_1A r\cos\theta\cos\phi}{(A^2 + r^2
-2A r\sin\theta\cos\phi)^{\frac{3}{2}}} ~,\ \ee
\eqa \pa{\Phi '}{\phi} = \pa{\Phi}{\phi} - \nonumber\eea\eq\hfill 
\frac{GM_1r\sin\theta\sin\phi}{A^2} - 
\frac{GM_1A r\sin\theta\sin\phi}{(A^2 + r^2
-2A r\sin\theta\cos\phi)^{\frac{3}{2}}} ~.\ \ee

\subsection{Accuracy}
In Appendix~\ref{implementation}, we describe the implementation of
this potential calculation in the code.
To test the accuracy of the potential calculation, we compared it to
three simple cases in which analytical solutions exist. We also
compared it to the \zeus~code (Stone \& Norman 1992) for which the
same tests have been performed. For the purposes of comparison, we 
chose their test cases in spherical polar coordinates in which 
they assumed equatorial symmetry as in our calculations.
They do not list the number of grid points they used. Here we used
$50^3$ grid points. Table~\ref{potcomp} shows the
comparison in the potential calculation for three cases of a
homogeneous sphere, a centrally condensed sphere and a homogeneous
ellipsoid.

\begin{table*} 
\begin{center}
\begin{tabular}{lcc}
\hline
Object & Max. error in $\pa{\Phi}{r}$ & Max. error in $\grad\Phi$ in \zeus \\
\hline
Homogeneous Sphere & 1.02$\times$10$^{-8}$\% & 1.11$\times$10$^{-2}$\% \\
Centrally Condensed Sphere & 6.35$\times$10$^{-2}$\% & 9.98$\times$10$^{-2}$\% \\
Homogeneous Ellipsoid & 1.53\% & 1.86\% \\
\hline
\end{tabular}
\end{center}
\caption{A comparison of the potential calculation in our code, 
performed on a three-dimensional grid, to the two-dimensional
calculations using the \zeus~code for three simple cases with analytic
solutions.}
\label{potcomp} 
\end{table*}

Table~\ref{potcomp} shows that our potential calculation is more
accurate than that of \zeus~regardless of the density distribution. As
our calculation is based on a numerical representaion of the
analytical value it is extremely accurate for the simplified case of a
homogeneous sphere. The centrally condensed sphere shows the accuracy
of the radial integration which is resolution limited.
The error in the calculation
for the homogeneous ellipsoid is smaller than that of the
\zeus~calculation. Indeed, it is only large at the surface --
throughout the remainder of the star it remains of similar order as in
the calculation for the centrally condensed sphere. We have also
calculated the error in the $\theta$ derivative of the potential. We
find that this is of similar order to the error in the $r$ direction.

\section{Solution method} \label{solmethod}
\subsection{Operator splitting of the equation of motion} \label{elmsect}
The equation of motion is solved in two parts using the method of
operator splitting or fractional step. With this technique the
equation of motion is split into two parts. One representing the
advective terms, and another representing the force terms. The
advantage of this method is that different numerical techniques may be
used to solve the two equations representing physically different
processes.

We split equation~(\ref{euler}) into two parts, one containing the
advective terms, the other the force terms. Once the first part is
solved, its solution is used in the second part to find the 
solution corresponding to the full time step.
Representing the time derivative of these equations in
finite difference form yields
\eq \frac{\ve{u}^{a} - \ve{u}^t_{\rm{p}}}{\delta t} +
2\ve{\Omega}\cross\ve{u}^{a} = 0 ~, \label{elmeq} \ee
 
\eq \frac{\ve{u}^{t+\delta t} - \ve{u}^{a}}{\delta t} =
-\frac{1}{\rho}\grad p - \grad\Phi '  + \ve{\mathcal{F}} ~, \ee
where we have followed the notation of Lu \& Wai (1998), and the
subscript p refers to the position of the element of interest in the previous
time step. Each part of the equation is solved for a full
time step ($\delta t$), and
we have used the notation $\ve{u}^a$ to
indicate the solution for the velocity once the first
equation, containing the advective terms, has been solved.

\begin{figure}
\center
\epsfig{file=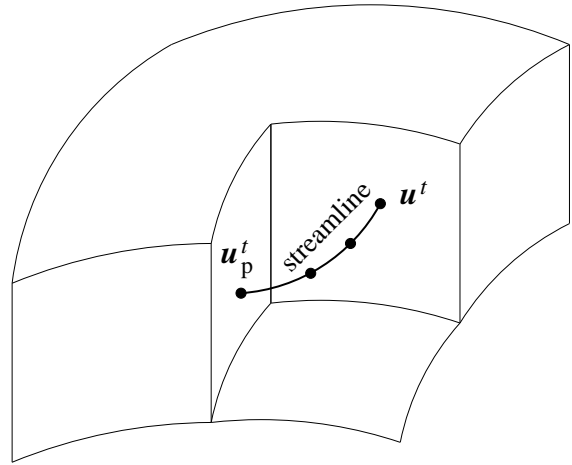,height=8cm,angle=-90}
\caption{Diagram showing the path interpolation used for the
Eulerian-Lagrangian method.} \label{elmfig}
\end{figure}

By solving the equations using the velocity of the fluid element
from the previous time step, the non-linear term
$(\ve{u} \dotp \grad )\ve{u}$ is eliminated from the equations. This means 
that advective terms in the equations are solved in a Lagrangian frame 
(for fixed mass elements) rather than an Eulerian frame (with fixed positions)
where the remainder of the terms are solved for. 
Hence this method is known as an Eulerian-Lagrangian method. 
Figure~\ref{elmfig} illustrates how
$\ve{u}_{\rm{p}}^t$ is calculated. The figure shows a cut away element
of the fluid. Using the velocity $\ve{u}^t$, we calculate the position
of the fluid element at a time step $\delta t$ before the present one.
Interpolating the velocity grid yields the velocity $\ve{u}_{\rm{p}}^t$ 
at this position. In actuality, we split the time step into a number 
of substeps enabling an accurate calculation of the streamlines of the
fluid. The substep method consists of finding the velocity at a point a
substep away and using this velocity to find the velocity in the next
substep. This way curved trajectories may be easily followed.

\subsection{Solution of the continuity equation on an adaptive grid}
The grid used in the code is adaptive. At the end of each time step, the
grid is rescaled so that its boundary represents the surface of the
star. Thus, the mass within each cell will change at the end of each
time step. To compensate for this in solving the integral form of the
continuity equation (equation~\ref{conuity}), the surface velocity is
calculated at each iteration of the continuity equation and the
radial velocities used in the calculation are relative to this
\eq u'_r(\theta ,\phi ) = u_r(\theta ,\phi ) - 
\frac {i}{\nr - N_{\rm c}} u_r(\theta ,\phi )\big |_{\rm{s}} ~, \ee
where the prime indicates the velocity used in the calculation and the
subscript s a quantity evaluated at the surface. Dorfi 
shows in LeVeque et al.\ (1998) [p.~279] that time centering 
of the equations results in
second order accuracy in $\delta t$. Hence the variables are evaluated at
half-odd-integer time steps i.e. when calculating the mass flow
through a cell boundary we use $\ve{u}\ts$ and
$\rho\ts$.

\subsection{Calculation of the pressure gradient at the surface}
The radial velocity needs to be calculated at the surface which
would require a ghost point for pressure half a radial grid zone past the
surface. Instead we use the surface
pressure which is one of the boundary conditions. Thus we
evaluate the pressure gradient one quarter of a zone below the
surface. However, this still yields an incorrect value for the
surface pressure gradient, so that in equilibrium it will not balance
the potential gradient causing a spurious surface velocity. To correct
for this, we calculate the potential gradient a quarter of a zone
below the surface and find that this accurately balances the pressure
gradient.

\subsection{Iteration procedure}
\begin{figure}
\begin{center}
\epsfig{file=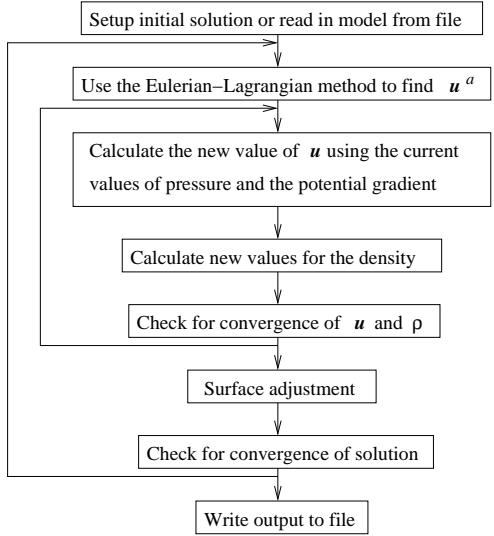,width=7.5cm,angle=-90}
\end{center}
\caption{A flow chart diagram showing the iteration procedure used in
the code} \label{flow}
\end{figure}

Figure~\ref{flow} shows the iteration procedure used in the
code. After initialization the main loop of the code is entered. The
Eulerian-Lagrangian method is used to find the solution of the
advective terms. A sub loop is then entered in which the velocities
and density are solved. Using the current values of pressure and density, the
new values for the velocities are calculated. With these velocities
the new densities can be determined. Once these have converged, the
end of a time step has been reached and the surface is adjusted to fit
the new shape of the star. Overall convergence is then tested 
until a solution is found which is written to an output file.
It is not guaranteed that the iteration procedure will always
converge.

\section{Advection tests} \label{advect}
To test the advection in our code, we use two simple tests. Both are
in spherical geometry and have also been
carried out by Stone \& Norman (1992) for their code \zeus~to which we
compare it.

\subsection{Expansion of a homogeneous sphere in a velocity field}
The first test is a ``relaxation'' problem for a pressure-free and
gravity-free homogeneous gas with a velocity field proportional to the
radius ($u_{\rm{r}} = v_0r$). The density will decay exponentially
with time as
\eq \rho (t) = \rho _0e^{-3v_0t} ~. \ee
At $v_0t=4$, the density should have decayed by almost six orders
of magnitude to $\rho=6.144\cross 10^{-6}\rho _0$. In our calculation
we use $v_0 = 1, \rho _0 = 1$ and three time steps of $10^{-2},10^{-3}$
and $10^{-4}$. The results are shown in Table~\ref{velfd}.

\begin{table} 
\begin{center}
\begin{tabular}{ccc}
\hline
Time step & Density at $t$ = 4 & Error in the density at $t$ = 4 \\
\hline
$10^{-2}$ & 6.522$\cross 10^{-6}$ & 6.14\% \\
$10^{-3}$ & 6.181$\cross 10^{-6}$ & 0.60\% \\
$10^{-4}$ & 6.148$\cross 10^{-6}$ & 0.06\% \\
\hline
\end{tabular}
\end{center}
\caption{A comparison of the calculated density to the analytical
solution for the expansion of a homogeneous sphere in a velocity
field.} \label{velfd} 
\end{table}

At $t=4$ our code is in better agreement than
\zeus~(5.60$\cross 10^{-6}$, 8.8\%) for all the time steps
considered, the error in the density being directly proportional to the
time step considered.  
Stone \& Norman (1992), however, provide no information on the time
step they use, so this cannot be compared further.

\subsection{Pressure-free collapse of a sphere under gravity}
The second advection test we use is the collapse of a homogeneous,
pressure-free sphere under gravity. Hunter (1962) showed that
\eq \frac{r}{r_0} = \cos ^2 \beta ~, \quad \frac{\rho}{\rho _0} = \cos
^{-6}\beta ~, \ee
where
\eq \beta + \frac{\sin 2\beta}{2} = t \sqrt{\frac{8\pi G\rho _0}{3}}
~, \ee
and $r_0$ and $\rho_0$ are the initial values of the radius and the
density, respectively. For $r_0 = \rho _0 = G = 1$, the free-fall time (the
time at which the sphere has collapsed to a point) is 0.543. The test
was run at two different time steps ($10^{-3}$ and
$10^{-4}$), and the density, radius and the mass were compared to the
analytical solution and the results obtained with \zeus.

\begin{figure}
\begin{center}
\epsfig{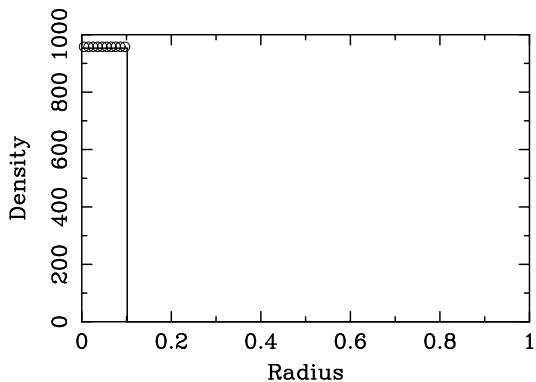}
\end{center}
\caption{Test simulation for the pressure-free collapse of a homogeneous
sphere under gravity. The comparison shows the density as a function
of radius at dimensionless time $t$ = 0.535
for a time step of $10^{-4}$. The solid curves represent the
analytical solution and the open circles our model results.}
\label{zeuscomp}
\end{figure}

\begin{table} 
\begin{center}
\begin{tabular}{cc}
\hline
Time step & Max. error in density at $t$ = 0.535 \\
\hline
$10^{-3}$ & 5.30\% \\
$10^{-4}$ & 0.50\% \\
\hline
\end{tabular}
\end{center}
\caption{A comparison of density calculated in the code to the
analytical solution for the collapse of a pressure-free sphere.}
\label{colpse} 
\end{table}
Table~\ref{colpse}~shows the errors in the calculated density in
comparison to the analytical solution. In all cases, the density
profile remained flat and the mass constant. This
demonstrates the ability of the code to adjust the surface of the star
correctly. Figure~\ref{zeuscomp}~shows a comparison at time $t$ =
0.535 with the analytical solution for a time step of $10^{-4}$. Note
how the grid is only defined within the sphere allowing an accurate
representation of the problem with no spurious points exterior to the
surface. In the calculation with \zeus, a number of grid points end up
outside the surface -- a direct consequence of their use of ghost zones
beyond the boundary.

\section{Comparison to Lai, Rasio \& Shapiro} \label{comparison}
\subsection{The compressible Maclaurin sequence} \label{lrsunits}
Lai, Rasio \& Shapiro (1993, hereafter LRS) calculated a sequence of
compressible Maclaurin spheroids based on
an energy variational method. 
LRS calculate universal dimensionless quantities
that are functions of the eccentricity of the spheroid
only.
The eccentricity is defined as
\eq e = \sqrt{1-\frac{R_{\rm{pole}}^2}{R_{\rm{equator}}^2}} ~, \ee
where $R_{\rm{pole}}$ and
$R_{\rm{equator}}$ are the radii of the pole and equator
respectively.
Transforming into the dimensionless units used in our code, their
equation (3.27) becomes
\eqa \hat{\Omega}^2 \equiv \kappa_n(1-n/5)\left ( \frac
{\tilde{\Omega}^{^2}\rhoc}{\pi\bar{\rho}} \right ) ~, \nonumber\eea
where
\eq \bar{\rho} = \frac{M}{4\pi\bar{R}^3} ~, \ee
is the mean density and
\eq \bar{R} = {(R_{\rm{pole}}^{} R_{\rm{equator}}^2 )^{\frac{1}{3}}}
~, \ee
is the mean value of the radius of the star.

For a polytropic system, as the polytropic index, $n$, increases, so
does the radius of the
polytrope (which is defined as the position where the density
drops to zero). For non-zero $n$ only one analytical solution exists
which has a finite radius. This is the case $n=1$ and has the solution
\eq \omega = \frac{\sin\xi}{\xi} ~, \ee
where
\eq \xi = r\sqrt{\frac{4\pi G}{(n+1)K}\rhoc^{1-\frac{1}{n}}}~, \quad \omega
= \left ( \frac{\rho}{\rhoc} \right )^{\frac{1}{n}} ~. \ee
This is one of the values for the polytropic index considered in both
the calculations of LRS and \uryu~(1998). Other values for the polytropic
index of interest are $n = 1.5,~3$. These polytropic indices are reasonable
representations for the internal structure of convective and radiative stars, 
respectively. Table~\ref{polytable} gives the ratio of mean
density to central density $(\bar{\rho}/\rhoc)$ for various polytropic
indices (Kippenhahn \& Weigert 1994).

\begin{table}
\begin{center}
\begin{tabular}{lcc}
\hline
$n$ & $\bar{\rho}/\rhoc$ \\
\hline
1 & 3.03963 $\cross 10^{-1}$ \\
1.5 & 1.66925 $\cross 10^{-1}$ \\
3 & 1.84561 $\cross 10^{-2}$ \\
5 & 0.0 \\
\hline
\end{tabular} 
\end{center}
\caption{Values of $\xi$ at the
surface of a polytrope and the ratio of mean density to central
density for various polytropic indices $n$.} 
\label{polytable}
\end{table}

In our tests we considered the case $n = 1$. This is the only
analytical solution which has a finite radius making it a
convenient starting approximation. 
Our code can also run, however, with approximate solutions appropriate
for $n = 1.5~\&~3$ using the method of Liu (1996) to calculate the
density distributions.
Starting from this analytic solution we
have calculated a compressible Maclaurin sequence for comparison, where 
we use 50 grid zones in the $r$ direction and 48 grid
zones in the $\theta$ direction. We increased the value of $\Omega$ in
units of 0.1 and used a time step of $\delta\tilde{t} = 0.001$ to find
a converged solution. We also included a linear artificial 
viscous pressure to damp out oscillations.

\begin{figure}
\begin{center}
\epsfig{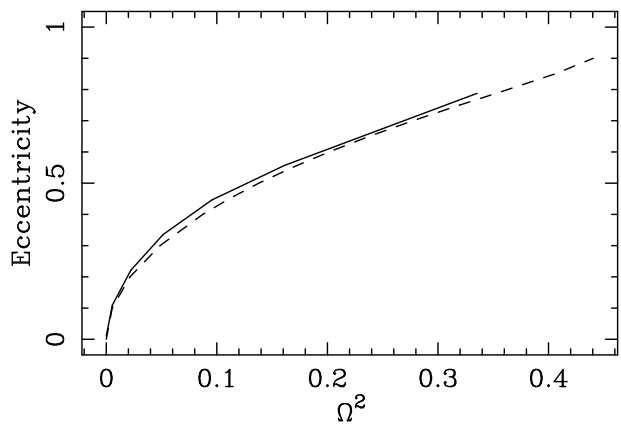}
\end{center}
\caption{Angular velocity squared against eccentricity of
an ellipsoid.  The dashed curve represents the results of LRS 
and the solid curve our results for comparison.}
\label{lrsfig}
\end{figure}

Figure~\ref{lrsfig} represents a
comparison to the calculations of LRS. It shows a
plot of the dimensionless angular velocity squared against
eccentricity. The figure shows that the code accurately calculates the
eccentricity for each value of $\Omega$.

\section{Future applications}
Our first application of the code will be the standard Roche 
problems. These have recently been considered by
\uryu~(1998), who assumed irrotational flows. We
will apply our code to the same case for comparison and then relax 
the constraint of irrotation, using realistic surface boundary conditions.
The second and main application will be the case of irradiation by X-rays
in interacting binaries, where we will study the effects of surface
heating, irradiation pressure and surface stresses to obtain
self-consistent solutions of the geometry and the circulation system
in irradiated systems and compare these results to observed systems
where irradiation effects have been identified (e.g. HZ Her/Her X-1,
Cyg X-2, Nova Sco, Vela X-1, Sco X-1). At a later stage we will 
extend the code to more realistic stellar models as outlined
in Appendix~\ref{generalisation}.

\bsp

\begin{appendix}
\onecolumn
\section{System variables}
Following are a list of variables used in this paper and a
description of what they represent: \\
\rule{0pt}{5pt}\\
\begin{tabular}{ll}
$\frac{d}{dt}$ & Lagrangian derivative \\
$\pa{}{t}$ \rule[0pt]{0pt}{12pt} & Eulerian derivative \\
$p$ & Pressure \\
$\rho$ & Density \\
$\ve{\mathcal{F}}$ & Frictional force \\
$\ve{u}$ & Velocity in the rotating frame \\
$\ve{u}^a$ & Velocity after the solution of the
advective terms in the equation of motion \\
& (section~\ref{elmsect}) \\
$\Omega$ & Angular velocity of rotation \\
$\Phi$ & Potential \\
$G$ & Gravitational constant \\
$g$ & Surface gravity \\
$r,~\theta ,~\phi$ & Spherical Polar coordinates \\
$A$ & Separation of the binary \\
$r_2$ & Distance from the primary to the point $(r,~\theta ,~\phi)$ \\
$M_1$ & Mass of the primary \\
$M_2$ & Mass of the secondary \\
$\ve{\hat{n}}$ & Unit surface normal vector \\
$\ve{\hat{r}},~\ve{\hat{\theta}},~\ve{\hat{\phi}}$ & Unit vectors in the
spherical coordinate system \\
$K, n$ & Constants in the polytropic equation of state
(equation~\ref{polytrope}) \\
\~~ & Superscript denoting dimensionless variable (section
\ref{dimensionless}) \\
$\nr,~\nth,~\nph$ & Number of $r,~\theta ,~\phi$ elements \\
$N_{\rm c}$ & Number of innermost radial zones whose position is kept
fixed \\
$\rm{i,j,k}$ & Subscripts used to denote $r,~\theta ,~\phi$ elements \\
$\tau$ & Local optical depth in the secondary\\
$\kappa$ & Local value of the opacity \\
$L_{\rm{x}}$ & X-ray luminosity of the compact object \\
$q$ & Artificial viscous pressure \\
${\bf\sf Q}$ & Artificial viscous pressure tensor \\
$C_1$ & Constant of order unity used in the calculation of artificial
viscous pressure \\
$C_{\rm{a}}$ & The adiabatic speed of sound \\
$\xi ,~\omega$ & Variables in the Lane-Emden equation for a polytrope
(section~\ref{comparison}) \\
$V$ & Volume of the secondary \\
$I$ & Moment of inertia of the secondary \\
$J$ & Angular momentum of the system \\
$T$ & Kinetic energy of the secondary \\
$W$ & Potential energy of the secondary \\
\^~ & Superscript denoting universal dimensionless variable with
no dependence on \\
& polytropic index (section~\ref{comparison}) \\
$\bar{R}$ & Mean radius of the secondary \\
$R_{\rm{pole}}$ & Polar radius of an ellipsoid \\
$R_{\rm{equator}}$ & Equatorial radius of an ellipsoid \\
$R_0$ & Radius of a spherical polytrope of mass equal to the secondary
\end{tabular}

\section{Generalisation to a non-polytrope} \label{generalisation}
In this appendix we describe how the computer code may
be extended to more realistic stellar models. The most important
change is to use a realistic (non-polytropic) equation of state
and the addition of an energy equation, the thermodynamic 
equation (Tassoul 1978):
\eq \rho \frac{d U}{d t} + p \grad  \dotp \ve{u} = \Phi_{\nu} + \rho
\epsilon_{\rm{nuc}} - \grad \dotp (\ve{F}+\ve{F_{\rm{r}}}) ~, \ee
where $U$ is the total internal energy per unit mass, $\Phi_{\nu}$ 
the heat generation by viscous friction, $\epsilon_{\rm{nuc}}$  the
rate of energy released by thermonuclear reactions per unit mass,
$\ve{F}$  the heat flux,
\eq \ve{F} =-\chi\grad T ~, \ee
and $\ve{F_{\rm{r}}}$ is the radiative flux,
\eq \ve{F_{\rm{r}}} = -\frac{4a c T^3}{3\kappa\rho} \grad T ~, \ee
where $a$ is the radiation constant and $T$  the temperature. The
thermodynamic equation may also be written as
\eq \rho T \frac{d S}{d t} = \Phi_{\nu} + \rho
\epsilon_{\rm{nuc}} - \grad \dotp (\ve{F}+\ve{F_{\rm{r}}}) ~, \ee
where $S$ is the entropy per unit mass. Landau \& Lifshitz (1987)
show that 
\eq \pa{S}{t} = \left (\pa{S}{t} \right )_p \pa{T}{t}, \quad \grad S =
\left (\pa{S}{t} \right )_p \grad T ~. \ee
Therefore,
\eq T \frac{d S}{d t} = c_{\rm{p}} \frac{d T}{d t} ~, \ee
where $c_{\rm{p}}$ is the specific heat at constant pressure. This
yields
\eq \rho c_{\rm{p}} \frac{d T}{d t} = \Phi_{\nu} + \rho
\epsilon_{\rm{nuc}} - \grad \dotp (\ve{F}+\ve{F_{\rm{r}}}) ~. \ee
To relate pressure, density and temperature, we need an equation of
state for the fluid, the simplest of which is given by the ideal gas law,
\eq p = \frac{k_{\rm{B}}}{\mu m_{\rm{H}}} \rho T ~, \ee
where $k_{\rm{B}}$ is Boltzmann's constant, $\mu$ the mean molecular
weight and $m_{\rm{H}}$ the mass of the hydrogen atom.

The energy generation rate per unit mass, and the opacity are often 
represented as power laws of temperature and density
(\uryu~1995), i.e.
\eq \epsilon_{\rm{nuc}} = \epsilon_0 \rho^{\gamma} T^{\delta} ~, \ee

\eq \kappa = \kappa_0 \rho^{\alpha} T^{-\beta} ~, \ee
where $\epsilon_0$, $\gamma$, $\delta$, $\kappa_0$, $\alpha$ and
$\beta$ are constants. For simplified models for lower main-sequence
stars, appropriate values for these constants are
\eq \gamma \simeq 1 ~, \quad \delta \simeq 4.5 ~, \ee
and for Kramers' opacity law,
\eq \alpha = 1 ~, \quad \beta = 3.5 ~. \ee
Inclusion of the thermodynamic equation introduces temperature as
another variable into the problem along with other variables which
depend on $p,\rho,\ve{u}$ and $T$. This leads altogether to six
equations and six unknowns. It also requires an additional surface
boundary condition, which can be determined from an atmosphere
calculation where the effects of external irradiation (if present) may
also be included, just as in the case for standard stellar-structure
calculations (e.g. Kippenhahn, Weigert \& Hofmeister 1967). These
equations have to be solved simultaneously to yield the solution at
each time step.

\section{The calculation of the potential}
In this appendix we derive analytical formulae for the derivatives
of the gradient of the potential and describe the technique used
in their implementation.

\subsection{The derivatives of the potential}\label{dplmdth}

Differentiating equations~(\ref{mullereq}) to (\ref{dl}) with
respect to $r$, we obtain:
\eq
\pa{\Phi(r, \theta , \phi )}{r} = -G \sum_{l=0}^{\infty}
\frac{4\pi}{2l+1} \sum_{m=-l}^l Y^{l m}(\theta , \phi)\bigg
[\frac{-(l+1)}{r^{l+2}}\clm + 
\frac{1}{r^{l+1}}\pa{\clm}{r} + 
l r^{l-1}\dlm + r^l\pa{\dlm}{r}
\bigg ] ~.
\ee
\eq
\pa{\clm}{r} = r^{l+2} \int_{4\pi} d\Omega ' Y^{l m*}(\theta ', \phi ') 
\rho(r', \theta ', \phi ') ~, \ee

\eq \pa{\dlm}{r} = -r^{1-l} \int_{4\pi} d\Omega ' Y^{l m*}(\theta ', \phi ') 
\rho(r', \theta ', \phi ') ~.
\ee
Hence,
\eq \frac{1}{r^{l+1}}\pa{\clm}{r} + r^l\pa{\dlm}{r} = 0 ~, \ee
and the $r$ derivative of the potential becomes
\eq \pa{\Phi(r, \theta , \phi )}{r} = -G \sum_{l=0}^{\infty}
\frac{4\pi}{2l+1} \sum_{m=-l}^l Y^{l m}(\theta , \phi)\bigg
[\frac{-(l+1)}{r^{l+2}}\clm + l r^{l-1}\dlm \bigg ] ~.
\ee
The $\theta$ derivative is given by
\eq \pa{\Phi(r, \theta , \phi )}{\theta} = -G \sum_{l=0}^{\infty}
\frac{4\pi}{2l+1} \sum_{m=-l}^l \pa{Y^{l m}(\theta , \phi )}{\theta}
\left [ \frac{1}{r^{l+1}}\clm+r^l\dlm \right ] ~. \ee
Using the definition of the spherical harmonics, we find
\eqa
\pa{\Phi(r, \theta , \phi )}{\theta} = -G \sum_{l=0}^{\infty}
\frac{4\pi}{2l+1} \sum_{m=-l}^l \frac{Y^{l m}(\theta , \phi
)}{P_l^{|m|}(\cos \theta)} \frac{d\left [ \plm(\cos \theta )
\right ]}{d\theta} \left [ \frac{1}{r^{l+1}}\clm+r^l\dlm \right ] ~.
\eea
Using one of the recurrence relations for Legendre
polynomials,
\eq \plm(\cos \theta ) = (-1)^m \sin ^m \theta \frac{d^m \left [
P_l(\cos\theta ) \right ] }{d(\cos\theta )^m} ~, \ee
we then obtain
\eqa
\plmp(\cos \theta ) & = & (-1)^{m+1} \sin ^{m+1} \theta \frac{d^{m+1} \left [
P_l(\cos\theta ) \right ] }{d(\cos\theta )^{m+1}} ~, \nonumber\\
& = & (-1)^{m+1} \sin ^{m+1} \theta \left (
\frac{-1}{\sin \theta} \right ) \frac{d}{d\theta} \left [
\frac{\plm(\cos \theta )}{(-1)^m \sin ^m \theta} \right ] ~, \nonumber\\
& = &\frac{d\left [ \plm(\cos\theta ) \right ] }{d\theta} -
\plm(\cos\theta )\left ( \frac{m\cos\theta\sin
^{m-1}\theta}{sin^m\theta} \right ) ~.
\eea
Rearranging these, we get the required relation
\eq \frac{d\left [ \plm(\cos \theta )\right ]}{d\theta} =
\plmp(\cos\theta ) + \frac{m\cos\theta}{\sin\theta}\plm(\cos\theta )
~. \ee
When $m=l$, we set $\plmp$ equal to zero. The $\theta$ derivative is then
\eq
\pa{\Phi(r, \theta , \phi )}{\theta} = -G \sum_{l=0}^{\infty}
\frac{4\pi}{2l+1} \sum_{m=-l}^l Y^{l m}(\theta , \phi) \left [
\frac{P_l^{|m+1|}(\cos \theta)}{P_l^{|m|}(\cos \theta )} +
\frac{m\cos\theta}{\sin\theta} \right ] 
\left [ \frac{1}{r^{l+1}}\clm+r^l\dlm \right ]
~. \ee
Finally, the $\phi$ derivative is simply given by
\eqa \pa{\Phi(r, \theta , \phi )}{\phi} & = & -G \sum_{l=0}^{\infty}
\frac{4\pi}{2l+1} \sum_{m=-l}^l \pa{Y^{l m}(\theta , \phi )}{\phi}
\left [ \frac{1}{r^{l+1}}\clm+r^l\dlm \right ] ~, \nonumber\\ 
& = & -G \sum_{l=0}^{\infty}
\frac{4\pi}{2l+1} \sum_{m=-l}^l i m Y^{l m}(\theta , \phi )\left [
\frac{1}{r^{l+1}}\clm+r^l\dlm \right ] ~. 
\eea

\subsection{Implementation}\label{implementation}
To implement the expansion of the potential in spherical harmonics,
we follow the technique described in \muller~\& Steinmetz
(1995), in which the integrals are split into a sum of integrals over
sub-intervals. If we denote the position at which
equations~(\ref{cl})~\&~(\ref{dl}) are to be evaluated as $r_{\rm{n}}$, 
$C^{lm}(r_{\rm{n}})$ as $C^{lm}_{\rm{n}}$, and $D^{lm}(r_{\rm{n}})$ as
$D^{lm}_{\rm{n}}$:
\eq C^{l m}_{\rm{n}} = \sum_{\rm{j=1}}^{\nth}\sum_{\rm{k=1}}^{\nph}\left [
\int_{\phi _{\rm{k-1}}}^{\phk}\int_{\theta _{\rm{j-1}}}^{\thj}
\sin\theta d\theta d\phi Y^{l m*}(\theta , \phi ) \sum_{\rm{i=1}}^n
\int_{r_{\rm{i-1}}}^{\ri} d r r^{l+2}\rho(r, \theta , \phi ) \right ]
~, \ee

\eq D^{l m}_{\rm{n}} = \sum_{\rm{j=1}}^{\nth}\sum_{\rm{k=1}}^{\nph}\left [
\int_{\phi _{\rm{k-1}}}^{\phk}\int_{\theta _{\rm{j-1}}}^{\thj}
\sin\theta d\theta d\phi Y^{l m*}(\theta , \phi ) \sum_{\rm{i=n+1}}^{\nr}
\int_{r_{\rm{i-1}}}^{\ri} d r r^{1-l}\rho(r, \theta , \phi ) \right ] ~.
\ee
Introducing $A^{lm}\ijk$ and $B^{lm}\ijk$,
\eq A^{l m}\ijk = \int_{\phi _{\rm{k-1}}}^{\phk} \int_{\theta
_{\rm{j-1}}}^{\thj} \int_{r_{\rm{i-1}}}^{\ri} \sin\theta d\theta d\phi d r
r^{l+2} Y^{l m*}(\theta , \phi ) \rho(r, \theta , \phi ) ~, \ee

\eq B^{l m}\ijk = \int_{\phi _{\rm{k-1}}}^{\phk} \int_{\theta
_{\rm{j-1}}}^{\thj} \int_{r_{\rm{i-1}}}^{\ri} \sin\theta d\theta d\phi d r
r^{1-l} Y^{l m*}(\theta , \phi ) \rho(r, \theta , \phi ) ~, \ee
we may write
\eq C^{l m}_{\rm{n}} = \sum_{\rm{i=1}}^n \sum_{\rm{j=1}}^{\nth}
\sum_{\rm{k=1}}^{\nph} A^{l m}\ijk = C^{l m}_{n-1} + \sum_{\rm{j=1}}^{\nth}
\sum_{\rm{k=1}}^{\nph} A^{l m}_{\rm{n,j,k}} ~, \ee

\eq D^{l m}_{\rm{n}} = \sum_{\rm{i=n+1}}^{\nr} \sum_{\rm{j=1}}^{\nth}
\sum_{\rm{k=1}}^{\nph} B^{l m}\ijk = D^{l m}_{n+1} + \sum_{\rm{j=1}}^{\nth}
\sum_{\rm{k=1}}^{\nph} B^{l m}_{\rm{n+1,j,k}} ~. \ee
We implement this as
\eq A^{l m}\ijk = \delta\phi \int_{\theta_{\rm{j-1}}}^{\theta_{\rm{j}}}
\sin\theta 
Y^{l m*}(\theta, \phk) d\theta \int_{r_{\rm{i-1}}}^{\ri} \rho(r, \thj, \phk)
r^{l+2} d r ~, \ee

\eq B^{l m}\ijk = \delta\phi \int_{\theta_{\rm{j-1}}}^{\theta_{\rm{j}}}
\sin\theta Y^{l m*}(\theta, \phk) d\theta 
\int_{r_{\rm{i-1}}}^{\ri} \rho(r, \thj, \phk)
r^{1-l} d r ~. \ee
For the $\theta$ integration we use a Romberg method. A numerical
integration is required due to the large $\theta$ size of the cells
near the poles which is a result of the equal spacing in $\cos\theta$.
For the radial integration we assume that the density varies linearly
with $r$ within each cell
\eqa
\rho(r, \thj, \phk) & = & \rho(r_{\rm{i-1}}, \thj, \phk) +
\frac{(r - r_{\rm{i-1}})\delta\rho}{(\ri - r_{\rm{i-1}})} ~, \\
& = &\rho_{\rm{i-1,j,k}} + \frac{(r - r_{\rm{i-1}})\delta\rho}{(\ri -
r_{\rm{i-1}})} ~,
\eea
where
\eq \delta\rho = \rho\ijk - \rho_{\rm{i-1,j,k}} ~. \ee
Hence
\eqa
A^{l m}\ijk = \delta\phi \int_{\theta_{\rm{j-1}}}^{\theta_{\rm{j}}}
\sin\theta d\theta 
Y^{l m*}(\theta, \phk) \left [ \left ( \rho_{\rm{i-1,j,k}} -
\frac{r_{i-1}\delta\rho}{\ri - r_{\rm{i-1}}} \right )
\frac{r^{l+3}}{l+3} + \frac{\delta\rho}{(\ri - r_{\rm{i-1}})}
\frac{r^{l+4}}{l+4} \right ]^{\ri}_{\ri-1} ~,
\eea

\eqa
B^{l m}\ijk = \delta\phi \int_{\theta_{\rm{j-1}}}^{\theta_{\rm{j}}}
\sin\theta d\theta 
Y^{l m*}(\theta, \phk) 
\left\{
\begin{array}{ll}
\left [ \left ( \rho_{\rm{i-1,j,k}} -
\frac{r_{\rm{i-1}}\delta\rho}{\ri
- r_{\rm{i-1}}} \right ) \ln r  + \frac{\delta\rho}{(\ri -
r_{\rm{i-1}})} \frac{r^{3-l}}{3-l}
\right ]^{\ri}_{\ri-1} & \quad {\rm if}~ l = 2 ~, \\
\left [ \left ( \rho_{\rm{i-1,j,k}} -
\frac{r_{\rm{i-1}}\delta\rho}{\ri
- r_{\rm{i-1}}} \right )\frac{r^{2-l}}{2-l}  + \frac{\delta\rho}{(\ri -
r_{\rm{i-1}})} \ln r
\right ]^{\ri}_{\ri-1} & \quad {\rm if}~ l = 3 ~, \\
\left [ \left ( \rho_{\rm{i-1,j,k}} -
\frac{r_{\rm{i-1}}\delta\rho}{\ri
- r_{\rm{i-1}}} \right )\frac{r^{2-l}}{2-l}  + \frac{\delta\rho}{(\ri -
r_{\rm{i-1}})} \frac{r^{3-l}}{3-l}
\right ]^{\ri}_{\ri-1} & \quad {\rm otherwise.}
\end{array}
\right.\hfill
\eea
In our calculations we consider spherical harmonics up to and
including the $l=14$ term. This choice follows from the results
presented by \muller~\& Steinmetz (1995) who find (as we do) that the
inclusion of terms of higher order does not significantly affect the
calculation for geometries in which the use of spherical coordinates
is appropriate.

\subsection{Calculation of the surface derivatives} \label{surfderiv}
We use the potential to calculate the surface derivatives by
considering an incremental change in the potential $d\Phi$
\eq d\Phi = \pa{\Phi}{r} dr + \pa{\Phi}{\theta} d\theta +
\pa{\Phi}{\phi}d\phi ~. \ee
It is simple to show that
\eq \pa{R}{\theta} = { \frac{ \frac{d\Phi_{\rm{s}}}{d\theta} -
\left.\pa{\Phi}{\theta}\right|_{\rm{R}}}
{\left.\pa{\Phi}{r}\right|_{\rm{R}}} } ~,\quad \pa{R}{\phi} = {\frac{  
\frac{d\Phi_{\rm{s}}}{d\phi} - \left.\pa{\Phi}{\phi}\right|_{\rm{R}}}
{\left.\pa{\Phi}{r}\right|_{\rm{R}} } } ~, \ee
where $\Phi_{\rm{s}}$ denotes the surface potential and all terms are
evaluated at the surface. When the surface
is an equipotential the first term in the numerator is zero.

\section{Finite difference form of equations}
In this appendix we give the finite difference form for the terms used
in equations~(\ref{euler})~\&~(\ref{conuity}). As velocities are
calculated on cell boundaries, and pressure and density at the center
of cells, it is simple to represent the pressure term
(from equation~\ref{euler}) in finite difference form as
\eq \frac{1}{\rho}\pa{p}{r} = \frac{2\nr (p_{\rm{i+1,j,k}} - p\ijk)}
{\rjk(\rho_{\rm{i+1,j,k}}+\rho\ijk)} ~, \ee
and similarly for the derivatives in the $\theta$ and $\phi$ directions.
The Coriolis term in the equation of motion is solved using
equation~(\ref{elmeq}). With the notation $u,v,w$ to represent
$u_{\rm{r}},u_{\theta},u_{\phi}$ for ease of reading, the three
components of equation (\ref{elmeq}) are
\eq \frac{u\ijk^a - u_{\rm{p}\ijk}^t}{\delta t} = 2\Omega
w_{\rm{i+\half,j,k+\half}}^a \sin\theta_j ~, \ee

\eq \frac{v\ijk^a - v_{\rm{p}\ijk}^t}{\delta t} = 2\Omega
w_{\rm{i,j-\half,k+\half}}^a \cos\theta_{j-\half} ~, \ee

\eq \frac{w\ijk^a - w_{\rm{p}\ijk}^t}{\delta t} = -2\Omega
\left ( u_{\rm{i-\half,j,k-\half}}^a \sin\theta_j +
v_{\rm{i,j+\half,k-\half}}^a \cos\theta_j \right ) ~. \ee
The integral form of the continuity equation (\ref{conuity}) in finite
difference form is
\eqa
\frac{\rho^{t+\delta t}\ijk - \rho^t\ijk}{\delta t} \delta \left (
\frac{r^3}{3} \right ) \delta (-\cos \theta) \delta\phi +
\rho\ts_{\rm{i+\half,j,k}}\big (
u\ijk\ve{\hat{n}} \dotp \ve{\hat{r}} +
v_{\rm{i+\half,j+\half,k}}\ve{\hat{n}} \dotp \ve{\hat{\theta}} +
w_{\rm{i+\half,j,k+\half}}\ve{\hat{n}} \dotp \ve{\hat{\phi}} \big )
r_{\rm{i+\half,j,k}}^2\delta\phi \delta (-\cos\theta ) -
\nonumber\eea\eqa\phantom{empty}\hfill
\rho\ts_{\rm{i-\half,j,k}}\cross \big (
u_{\rm{i-1,j,k}}\ve{\hat{n}} \dotp \ve{\hat{r}} +
v_{\rm{i-\half,j+\half,k}}\ve{\hat{n}} \dotp \ve{\hat{\theta}} +
w_{\rm{i-\half,j,k+\half}}\ve{\hat{n}} \dotp \ve{\hat{\phi}} \big )
r_{\rm{i-\half,j,k}}^2\delta\phi \delta (-\cos\theta ) +
\rho\ts_{\rm{i,j+\half,k}} v_{\rm{i,j+1,k}} \sin\theta_{j+\half}
r_{\rm{i,j+\half,k}} \delta r \delta\phi
\nonumber\eea\eq\phantom{empty}\hfill
-\rho\ts_{\rm{i,j-\half,k}} v\ijk \sin\theta_{j-\half}
r_{\rm{i,j-\half,k}} \delta r \delta\phi + 
\rho\ts_{\rm{i,j,k+\half}} w_{\rm{i,j,k+1}}
r_{\rm{i,j,k+\half}} \delta r \delta (-\cos\theta ) -
\rho\ts_{\rm{i,j,k-\half}} w\ijk
r_{\rm{i,j,k-\half}} \delta r \delta (-\cos\theta ) = 0 ~.\ \
\ee

\end{appendix}
\label{lastpage}
\end{document}